\documentclass[a4paper]{revtex4-1}
\usepackage{graphics}
\usepackage[english]{babel}
\usepackage[a4paper,left=3cm,top=3cm,right=2cm,bottom=3.5cm]{geometry}
\usepackage{amsmath}
\usepackage{textcomp}
\usepackage{xcolor}     
\usepackage{soul}

\begin{document}

\begin{large}
\noindent{\textbf{Quantum Phase Slips: from condensed matter to ultracold quantum gases\\}}
\end{large}

\noindent{C. D'Errico$^{1,2,*}$, S. Scaffidi Abbate$^{2}$ and G. Modugno$^{1,2}$}

\begin{small}
\noindent{$^{1}$Istituto Nazionale di Ottica, CNR, 50019 Sesto Fiorentino, Italy\\$^{2}$LENS and Dipartimento di Fisica e Astronomia, Universit\`{a} di Firenze, 50019 Sesto Fiorentino, Italy\\$^{*}$derrico@lens.unifi.it\\}
\end{small}

\textbf{Abstract:}
Quantum phase slips are the primary excitations in one-dimensional superfluids and superconductors at low temperatures. They have been well characterized in most condensed-matter systems, and signatures of their existence has been recently observed in superfluids based on quantum gases too. In this review we briefly summarize the main results obtained on the investigation of phase slips from superconductors to quantum gases. In particular we focus our attention on recent experimental results of the dissipation in one-dimensional Bose superfluids flowing along a shallow periodic potential, which show signatures of quantum phase slips.

\section{Quantum phase slips in condensed matter}

The ability to carry charge or mass currents with zero dissipation is the hallmark of superconductivity and superfluidity. Superconductors and superfluids are characterized by a macroscopic wave function $\Psi(r)=|\Psi(r)|e^{i\phi(r)}$, the order parameter. The amplitude and phase coherence of the order parameter vanish at the transition temperature or at the critical current, where the system goes back to the normal phase.
However, also below the critical temperature and the critical current, the phase coherence of the system can be affected by the so-called phase slips, i.e. phase fluctuations of the order parameter, which induce a finite resistance and therefore lead to a destruction of persistent currents.


Phase slips have been historically predicted by Little \cite{Little} as thermally activated topological defects, but it is now assessed that they  may occur even at zero temperature, due to quantum tunneling events \cite{Giordano}. \\
A phase slip event, in fact, is an elementary excitation of the order parameter due to thermal or quantum fluctuations, corresponding to a local suppression of its amplitude and a simultaneous jump of the phase by 2$\pi$. As a consequence of the phase slip event, the superfluid metastable state, i. e. a local minimum of the Ginzburg-Landau free energy $F$ \cite{Landau} with velocity $v\propto\nabla\phi(x)$, decays into a state with lower velocity, since the phase has locally unwound \cite{Little}. A phase slip can be thermally activated (TAPS)\cite{Langer,McCumber,Luckens} when the temperature is higher than the free-energy barrier $\delta F$ between two neighbouring metastable states and the system may overcome the barrier via thermal fluctuations. The nucleation rate of TAPS follows the Arrhenius law $\Gamma\propto e^{-\delta F/k_BT}$ \cite{Langer,McCumber}, making these events extremely improbable for $T\lesssim\delta F/k_B$. In this latter regime, a second mechanism for the activation of phase slips becomes dominant: quantum tunneling below the free-energy barrier, triggered by quantum fluctuations. This second type of excitations has been called quantum phase slips (QPS) \cite{Giordano}. Since the role of both thermal and quantum fluctuations becomes increasingly important when reducing the dimensionality, these phenomena are particularly relevant for one-dimensional systems.

\begin{figure}[h]
\begin{center}
\resizebox{0.5\columnwidth}{!}{\includegraphics{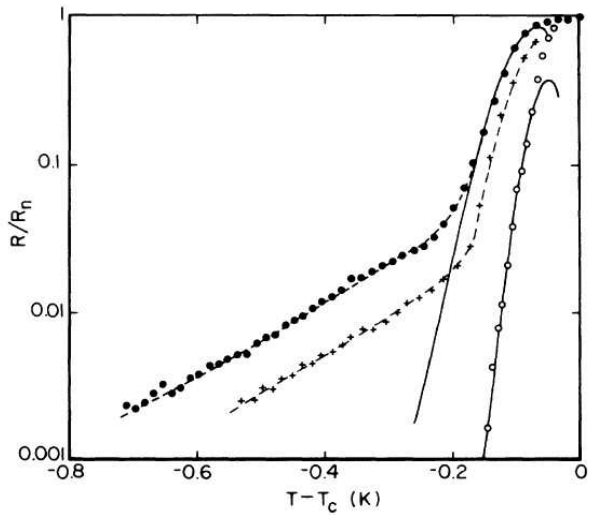}}
\caption{From thermal to quantum phase slips in nanowires. Resistance in In wires as a function of the temperature, for three different values of the wire diameter. Solid lines are the fits with a thermal activation mechanism, while dashed curves include both thermal and quantum contributions. Figure adapted from Ref. \cite{Giordano}.}
\label{figgiordano}
\end{center}
\end{figure}

QPS have been observed in different condensed-matter systems, such as superconducting nanowires \cite{Bezryadin01,Lau01,Altomare,Bezryadin09,Kamenev} and Josephson junction arrays \cite{Pop10}. In these systems it is typically possible to control the generation of QPS by changing the temperature or the current. Recently QPS have been proposed for applications such as topologically-protected qubits \cite{Mooij,Belkin} or as quantum standard for the electrical current \cite{Pop10}.\\
In 1967 William Little introduced the notion of phase slip due to TAPS \cite{Little} to describe the finite resistance below $T_C$ in thin wires made of superconductive material. At any non-zero temperature, in fact, each segment of a thin wire has a finite probability to become a normal conductor, for a very short time, due to phase-slip events. Each of these fluctuations could disrupt the persistent current and confer a non-zero resistance to the wire. This resistance, defined by an Arrhenius-type equation, drops exponentially with cooling but remains finite at any non-zero temperature as an effect of TAPS events. The height of the energy barrier $\delta F$ depends on several parameters, like the cross-sectional area of the wire, and decreases when increasing the applied current \cite{Little}. A more precise theory for TAPS was developed by Langer and Ambergaokar \cite{Langer}, and later completed by McCumber and Halperin \cite{McCumber}. Any thermal activation over a barrier is dominated by quantum tunneling at sufficiently low temperatures and phase slips should be induced by quantum fluctuation of the order parameter. The first observation of QPS was reported by Giordano \cite{Giordano}, in experiments with thin In and PbIn wires. By studying the resistance of the wires as a function of the temperature, Giordano observed a crossover from thermally activated behaviour near $T_C$ to a more weakly temperature-dependent resistance tail at lower temperatures (Fig. \ref{figgiordano}). In order to describe this observation, Giordano proposed a phenomenological description based on the Caldeira-Leggett model for macroscopic quantum tunneling of phase slips through the same free energy barrier expected for TAPS: the nucleation rate for QPS scales as $\Gamma\propto e^{-\frac{\delta F}{\hbar \omega_0}}$, where $\omega_0$ is an attempt frequency. As the collective state of a large number of electrons is involved, these phase slips are an example of macroscopic quantum tunneling, experimentally observed in Josephson junctions \cite{Martinis,Lee,Yu,Jackel} and superconducting loops \cite{Astafiev}. Many experiments on nanowires observed resistance tails and were discussed in terms of QPS \cite{Lau01,Giordano,Giordano&Schuler,Giordano90,Giordano91,Giordano94,Zgirski05,Altomare,Zgirski07}, but their detailed description goes beyond the scope of this article. For example, Ref. \cite{Bezryadin08} and Ref. \cite{Arutyunov08} give an accurate review of the suppression of superconductivity in 1D nanowires.

\section{Quantum phase slips in ultracold quantum gases}
\label{sec.coldatoms}

Thermal and quantum phase slips are relevant for all types of superfluids, including those based on ultracold quantum gases. The study of phase slips in quantum gases is appealing, since it could allow the investigation of QPS in regimes that are not accessible in condensed matter systems, thanks to the possibility of tuning key parameters such as dimensionality, density, interaction strength and temperature. In addition, the diluteness of quantum gases and the precise knowledge of their relevant physical parameters allow to perform accurate comparisons with theoretical models. During the last decade there has indeed been an increasing interest in studying phase-slip phenomena in various quantum gases configurations, both experimentally and theoretically \cite{Buchler,Polkovnikov02,Ruostekoski,Polkovnikov05,Khlebnikov,Danshita12,Danshita13,Roscilde,Kunimi}. An incontrovertible evidence of QPS has however not yet been obtained, although recent experiments have given strong indications of their presence.

One interesting configuration that is being investigated is the toroidal geometry, that is typically realized with specially shaped optical traps \cite{Moulder12, Wright13}. This configuration is not truly one-dimensional, since several transverse modes are present, so thermal and quantum fluctuations are low. These setups have been employed to study the decay of persistent currents \cite{Moulder12, Kumar17}, weak links, analogous to Josephson junctions \cite{Wright13}, and even more complex "atomtronic" configurations \cite{Jendrzejewski14}. Most of these studies have reported the observation of individual phase slips events at the critical velocity, or of resistive flow above $v_c$. The observation are in general well described by the same Feynman phase-slip model that has been applied also to superfluid He systems \cite{Savard11}. In one experiment with particularly long-lived supercurrents, also rare, stochastic phase-slip events have been observed at low velocities, well below $v_c$ \cite{Moulder12}. It is however still unclear whether they are of quantum or thermal nature.

Another configuration of interest is that of a narrow channel between two large reservoirs, which is being employed to study a variety of transport phenomena with fermionic \cite{Brantut12} and bosonic \cite{Eckel16} quantum gases. Also this type of setup is realized with optical potentials, and reaching the truly one-dimensional regime is hard because diffraction limits the transverse size of the channel. The setup has however been employed to study the transition between a resistive flow and a superflow, which is controlled by the flow velocity. Also in this configuration one can define a critical velocity above which it becomes energetically favorable to reduce the velocity by creating phase slips \cite{Eckel16}. Differently from the previous case, the phase slips can move along the channel and reach the reservoirs, where they decay into vortices and can be observed with the standard imaging techniques. A similar phenomenon has been recently observed also for fermionic superfluids in a Josephson-junction configuration \cite{Burchianti17}. These observations are well fit by the Feynman model for deterministic creation of phase slips, and below the critical velocity the superflow does not show apparent decay on the available time scale. Also in this configuration, therefore, it is apparently not possible to study the generation of the rare phase slips driven by quantum fluctuations.

\begin{figure}[b]
\begin{center}
\resizebox{0.5\columnwidth}{!}{\includegraphics{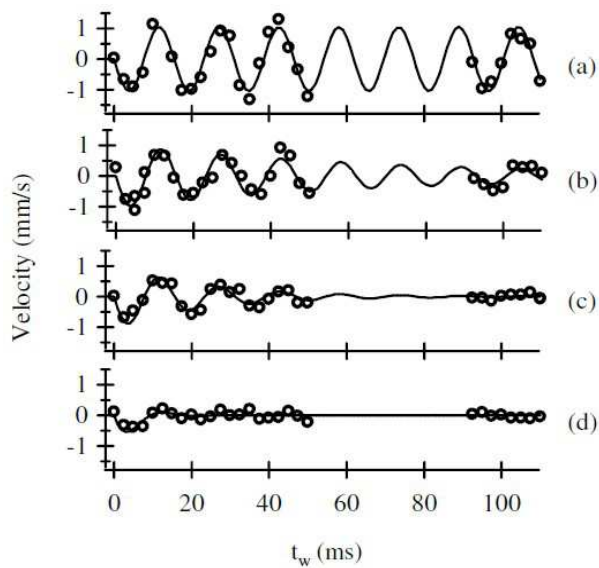}}
\caption{Increased role of thermal and quantum fluctuations in 1D quantum gases. Damped oscillations of a $^{87}$Rb 1D gas in an optical lattice, for different lattice depths, expressed in units of the atomic recoil energy, $s=V_0/(h^2/2m\lambda^2)$: (a) $s$ = 0, (b) $s$ = 0.25, (c) $s$ = 0.5, (d) $s$ = 2. Figure adapted from Ref. \cite{Fertig05}.}
\label{figporto}
\end{center}
\end{figure}

Finally, the transport in truly one dimensional systems is being studied using optical lattices, which employ interference to beat the diffraction limit. They are used to provide a strong confinement of a quantum gas along two spatial dimensions, reaching the regime in which the radial excitations are frozen and the only relevant dynamics happens along the third, longitudinal direction. This configuration has been often employed to study mass currents in the presence of a periodic potential along the longitudinal direction. Since a longitudinal harmonic potential is naturally present in this setup, the most common method to study the dissipation of the superflow is to excite a collective oscillation of the gas and to watch how it is damped in time \cite{Fertig05,Kasevich}. A more elaborate technique consists in moving the optical lattice relative to the harmonic potential \cite{Ketterle07}. From a general point of view, the periodic potential constitutes an obstacle to the superflow and therefore might lead to the excitation of phase slips. The potential does however modify more deeply the properties of the system, leading to a critical dynamical phenomenon at sufficiently large velocities, the dynamical instability, or to the Mott-insulator (MI) quantum phase transition, at sufficiently large atom-atom interactions and for commensurate lattice filling. The two types of instability are smoothly connected and it is possible to calculate a critical velocity $v_c$ that depends on the interaction strength and goes to zero at the MI transition \cite{Polkovnikov05}. In a 3D system, both dynamical instability and MI are characterized by sharp transitions, but in 1D the increased role of thermal and quantum fluctuations lead to a broadening of the transition, with a proliferation of phase slips even for subcritical velocities and interaction strengths \cite{Polkovnikov05}. Seminal experiments with bosonic superfluids have indeed observed a rather strong dissipation even for velocities much lower that $v_c$, which increase for increasing interaction strengths as shown in Fig. \ref{figporto}. However, they have not been able to identify the dominant mechanism for the dissipation in thermal or quantum phase slips \cite{Fertig05,Kasevich,Ketterle07}.

\begin{figure}[b]
\begin{center}
\resizebox{0.5\columnwidth}{!}{\includegraphics{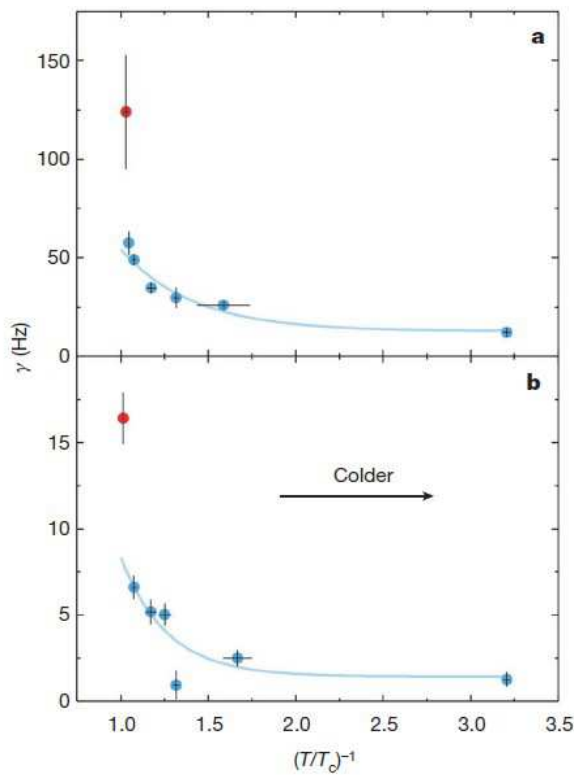}}
\caption{Onset of a temperature-independent phase slip regime. Temperature dependence of the damping rate $\gamma$, measured in oscillation of a $^{87}$Rb BEC in a 3D optical lattice for two values of the lattice depth: (a) $s$ = 6, (b) $s$ = 2. Figure adapted from Ref. \cite{Demarco08}. }
\label{figdemarco}
\end{center}
\end{figure}

A later study has explored the role of temperature, and also confirmed that the mechanism underlying the observed dissipation are phase slips, through the observation of vortices \cite{Demarco08}. In that study, the longitudinal lattice was so strong that the system was effectively prepared in a 3D lattice, and the velocity was kept in the range of velocities below $v_c$ where the dissipation is strong. As shown in Fig. \ref{figdemarco}, the damping rate of the oscillations showed a decrease with decreasing temperature, until a regime of approximately constant dissipation was reached at the lowest temperatures. A study of the formation of vortices in the sample confirmed that the dissipation was due to a phase-slip type of excitations. Such weak temperature dependence might suggest the onset of a regime where quantum phase slips dominate over thermal phase slips, but the experimentalists could not observe the very strong dependence of the dissipation rate on the interaction strength predicted for QPS \cite{Polkovnikov05}. A similar weak interaction dependence of the dissipation has been observed also for a periodic potential in 1D, for velocities close to $v_c$ \cite{Tanzi13}.

For 1D systems in periodic potentials, the theory predicts also a peculiarly difference of the dissipation rate due to thermal or quantum phase slips on the flow velocity \cite{Danshita13}. Exploring these phenomena for velocities close to $v_c$ is however impossible, due to the rapid variation of the dissipation rate with velocity and to the intrinsic variation of velocity in experiments based on an oscillating system. For these reasons, our group has recently started investigating the dissipation in 1D gases in the regime of very weak lattices and low velocities, which allow to access the limit $v \ll v_c$ \cite{Boeris} and to attempt a study of the velocity dependence of the dissipation rate \cite{Tanzi16,Scaffidi}. An important novelty of our approach has been the use of Bose gas with tunable interparticle interaction \cite{DErrico}, which allows to disentangle, for the first time, the interaction and tunneling energies in the system. As will be discussed in following, these studies apparently provide a strong indication of the onset of quantum phase slips in ultracold quantum gases, although the agreement with the theory is not yet satisfactory.

\section{Velocity-dependent dissipation in one-dimensional quantum gases}
This Section is devoted to a more detailed presentation of recent studies with one-dimensional ultracold bosonic gases in a weak periodic potential, which have explored for the first time the velocity dependence of the dissipation rate in various interaction regimes \cite{Boeris,Tanzi16,Scaffidi}. These studies have been motivated by the theoretical prediction of a peculiar velocity dependence of the QPS nucleation rate in the presence of periodic potentials \cite{Danshita13}. To avoid the diverging phase slip rates that occur close to the dynamical instability or the Mott insulator transition, the experiments have been carried out in the low velocity regime, after a careful characterization of those instabilities.

\subsection{Large oscillations: dynamical instability and Mott transition}
\label{subsec.largev}
The observation of the critical velocity for dynamical instability has been employed to determine the critical interaction strength for the onset of the Mott insulator both in the presence of a strong lattice \cite{Tanzi13} and in the presence of a weak one \cite{Boeris}.

\begin{figure}[b]
\begin{center}
\resizebox{0.8\columnwidth}{!}{\includegraphics{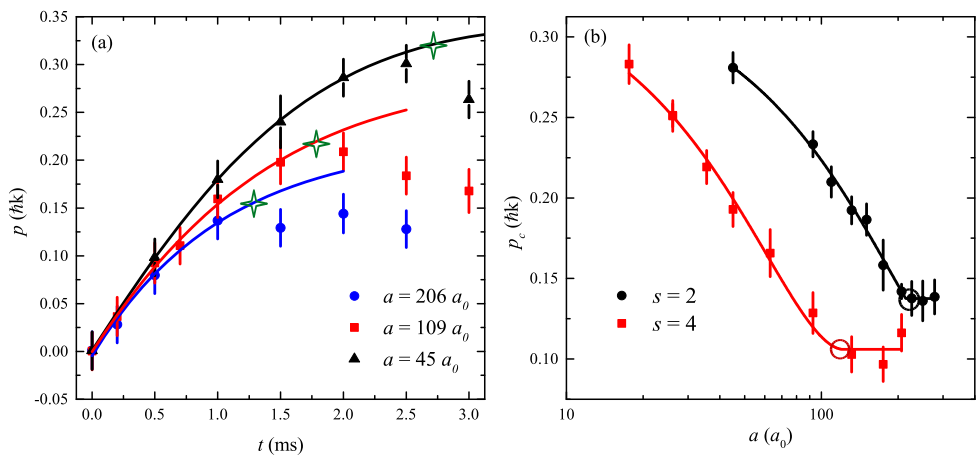}}
\caption{Detecting the dynamical instability and the Mott insulator transition. (a) Time evolution of the momentum distribution peak $p$ at $s$ = 2 for different values of scattering length. The solid lines are the theoretical damped oscillation fitting the data for $p<p_c$ before the dynamical instability sets in. The green stars mark the critical momentum $p_c$. (b) Critical momentum $p_c$ versus scattering length for two lattice depth: $s$ = 4 (red squares) and $s=2$ (black circles). A piecewise fit (solid lines) determines the critical values for the superfluid-Mott insulator transition (empty circles) for $n$ = 1.  Figure adapted from \cite{Scaffidi}.}
\label{figdi}
\end{center}
\end{figure}

Three examples of the time evolution of the position of the momentum distribution peak $p$, for a fixed lattice depth $s$ = 2 - in units of $^{39}$K recoil energy - and three different values of the scattering length, are shown in Fig. \ref{figdi}a. An initial increase of $p$ up to a certain critical value $p_c$ is followed by a subsequent decrease. This behavior can be analysed in the frame of a phase slips based model \cite{Tanzi13,Boeris,Polkovnikov05}. The observed dissipation, in fact, is induced by phase slips mechanism: at short times, for $p<p_c$, the oscillation is only weakly damped; at larger times the phase slip rate diverges and the system enters a dynamically unstable regime, characterized by a strong damping. The value of $p$ where the experimental data points deviate with respect to the theoretical curve can be identified with the critical momentum $p_c$ for the occurrence of the dynamical instability \cite{Tanzi13,Boeris}. By increasing the scattering length, the damping rate at short times increases, as the phase slips nucleation rate increases, while $p_c$ decreases.

At the superfluid-Mott insulator transition, the critical momentum for dynamical instability is expected to vanish. In Fig. \ref{figdi} is shown the behavior of $p_c$ as a function of $a$ for two different lattice depths. For increasing $a$, the measured $p_c$ initially decreases and then reaches a finite constant value.

The onset of the plateau can be interpreted as the critical scattering length $a_c$ to enter the Mott regime for the commensurate regions of the system. Despite the strong inhomogeneity of the systems, in fact, at sufficiently strong interactions the experiment observes a suppression of the system dynamics:  within each tube when a part of the atoms reaches the localization condition $n$ = 1, transport is suppressed also in the remaining adjacent parts with different occupation \cite{Boeris}.

For each set of measurements with a given value of $s$, $a_c$ can be determined by means of a second-order polynomial piecewise fit \cite{Boeris}, which is justified by the phase slip based model \cite{Danshita12,Polkovnikov05}. In the absence of an exact theoretical model to obtain the critical momentum to enter the dynamical instability regime at finite interaction, in fact, it is possible to use a quantum phase slips based model to predict the interaction dependence of $p_c$, which well reproduces the experimental behaviour of $p_c$ with the interaction and shows a quadratic dependence of $p_c$ on $a_c$ \cite{Boeris}. This result justifies the choice of the quadratic polynomial fit in Fig. \ref{figdi}b, where is shown that $a_c$ increases when $s$ decreases. Comparing experimental results, extracted for different values of the lattice depth, with a theoretical analysis based on quantum Monte Carlo simulations an excellent agreement has been found \cite{Boeris}. This investigation demonstrates that, also in the presence of a shallow lattice the onset of the Mott regime can be detected from a vanishing $p_c$, as in the presence of a deep lattice \cite{Tanzi13}.

\subsection{Small oscillations: velocity-dependent quantum phase slips}
\label{subsec.smallv}

For momenta lower than the critical momentum for dynamical instability, the system never enters in the unstable regime and keeps oscillating with the dissipation induced by phase slips nucleation. The damping rate $G$ is related to the phase slips nucleation rate $\Gamma$ via $G=\frac{h}{mL}\frac{\Gamma}{v}$ \cite{Danshita13,Tanzi16,Scaffidi}, where $L$ is the length of the chain. Theory predicts that the damping rate $G$, due to the presence of phase slips, exhibits different parameter dependence depending on the phase-slips activation mechanism. In particular, $G \simeq e^{\frac{-\delta F}{k_bT}}$ in the presence of TAPS, with $\delta F$ free energy barrier between the two metastable state, whereas $G \simeq v^\alpha$ for QPS and $G \simeq T^{\alpha - 1}$ in the intermediate case where QPS are assisted by temperature (TAQPS), with $\alpha$ being an interaction-dependent parameter \cite{Danshita13}. Below a characteristic temperature $T^*$ phase slip events are activated by quantum fluctuation and the damping rate is predicted to be temperature-independent, while above $T^*$ phase slips are assisted by temperature and the damping rate is velocity independent. The specific form of $\delta F$ and $T^*$ depends on the type of obstacle experienced by the system \cite{Danshita12,Khlebnikov,Buchler}. In the case of periodic potentials, the relevant energy scale is the Josephson plasma energy $E_j$ \cite{Danshita12,Danshita13}, which sets both the free-energy barrier $\delta F \simeq E_j$ and the crossover temperature $T^* \simeq E_j/k_B \times v/v_c$.

\begin{figure}[b]
\begin{center}
\resizebox{0.8\columnwidth}{!}{\includegraphics{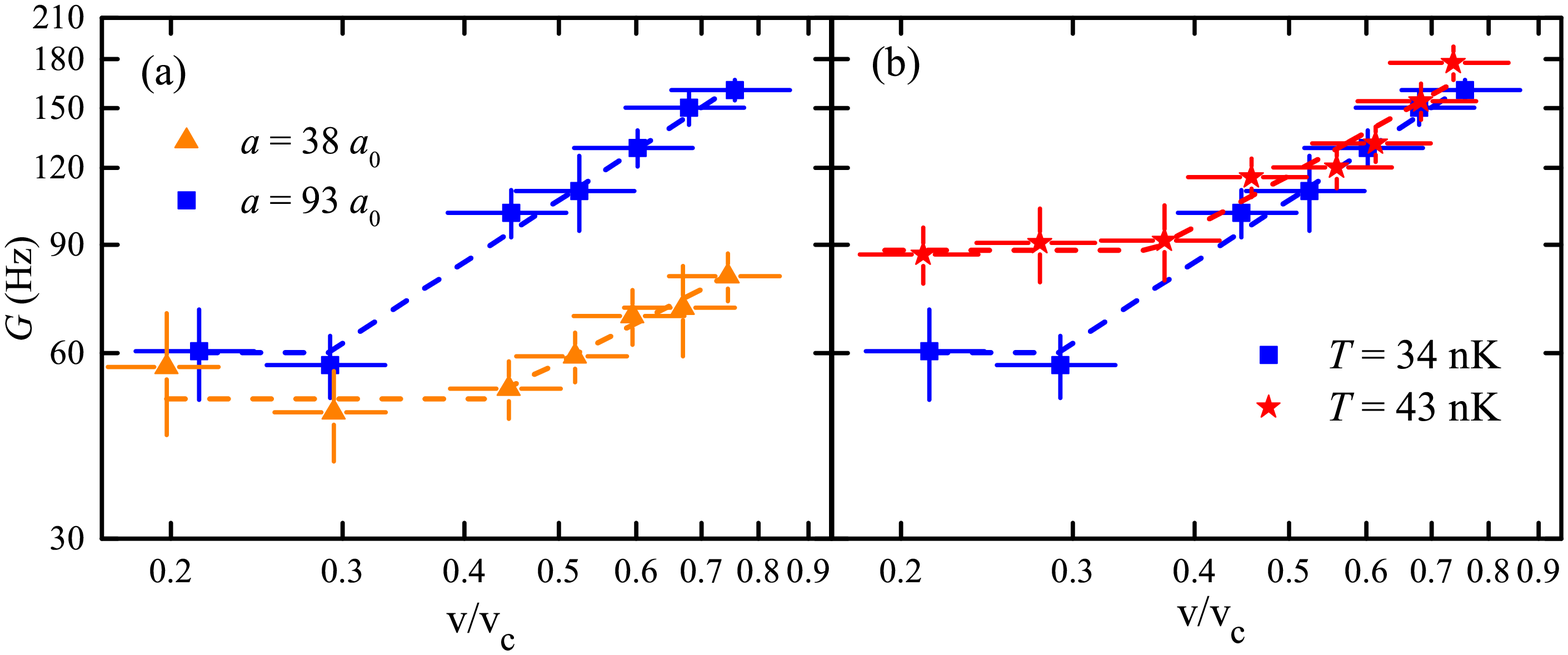}}
\caption{Onset of velocity-dependent dissipation. (a) Damping rate $G$ plotted as a function of the maximum velocity $v$ normalized to the critical velocity $v_c$, for two interaction strengths and constant temperature. (b) $G$ as a function of $v/v_c$ for two different temperatures and approximately constant interaction energy. The lines are fits to measure the crossover velocity $v^*$.  Figure adapted from \cite{Scaffidi}.}
\label{figps}
\end{center}
\end{figure}

The experimental observation of the crossover between TAQPS and pure QPS requires to measure the dissipation rate as a function of velocity. Nevertheless, in the regime of strong lattices ($s$ $\geq$ 5), where the theory predicts the linear dependence of $T^*$ with respect to $v$, the small critical velocity for dynamical instability strongly limits the range of accessible velocities. The first experimental attempt to the investigation of the TAQPS to QPS crossover has been performed in a shallow lattice ($s$ = 1). Fig. \ref{figps} shows the damping rate $G$ measured by fitting the oscillation of $p$ for a wide range of velocities $v$, interaction strengths and temperature. The velocity $v$, identified with the velocity reached during the first oscillation as in the theoretical model \cite{Danshita13}, has been rescaled to the corresponding critical velocity $v_c$, for different values of interaction at the same temperature (Fig. \ref{figps}a) or for different temperatures at the same interaction (Fig. \ref{figps}b). In both cases a crossover from a velocity-independent $G$ to a regime where $G$ grows with the velocity has been observed.

By fitting the datapoint with a piecewise linear function, the crossover velocity $v^*$, i. e. the minimum velocity required to enter the regime of dependence on $v$, is determined. The crossover velocity decreases for increasing interaction and increases for increasing temperature.
For $v\ll v^*$, the damping rate $G$ is strongly affected by temperature (Fig. \ref{figps}b), while the dependence on interaction is weaker (Fig. \ref{figps}a). Instead, in the $v$-dependent regime $G$ is dominated by interaction effects (Fig. \ref{figps}a) and a clear dependence on $T$ cannot be measured (Fig. \ref{figps}b). These observations appear consistent with the predicted crossover from thermally assisted to pure quantum phase slips, which can be controlled by changing the crossover temperature $T^*\propto E_j/k_Bv/v_c$ by tuning the velocity and the interaction strength. For $T^*\ll T$, i.e. at small velocity and small interaction, $G$ depends only on $T$ and it is velocity independent, suggesting a thermal activation of phase slips. For $T^*\gg T$, i.e. at large velocity and large interaction, the system enters in a regime where $G$ is linearly dependent on the velocity and temperature independent, suggesting a quantum activation of phase slips.

\begin{figure}[t]
\begin{center}
\resizebox{0.46\columnwidth}{!}{\includegraphics{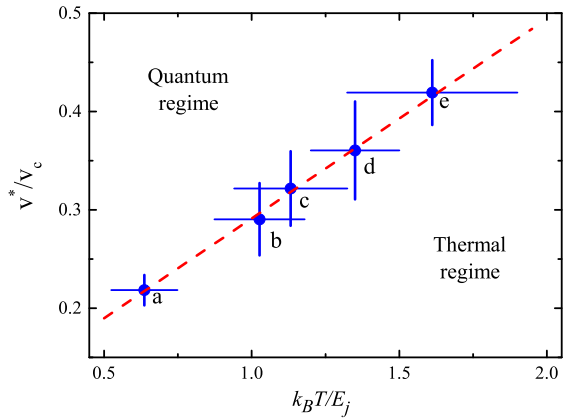}}
\caption{Experimental diagram of thermal and quantum dissipation regimes. Crossover velocity $v^*/v_c$ as a function of $k_BT/E_j$. The individual datapoints have been taken for different temperatures and interaction energies. The dashed line apparently separates the thermal and the quantum regimes for phase slips. Figure adapted from \cite{Scaffidi}.}
\label{figpd}
\end{center}
\end{figure}

Furthermore, Fig. \ref{figpd} shows the linear scaling of the crossover velocity $v^*/v_c$ as a function of temperature normalized to the Josephson energy, $k_BT/E_j$; from the fit we get $k_BT^*=4.9(14)E_jv/v_c-0.4(4)E_j$. This could be a further indication that the observed measurements are in agreement with the crossover from thermally assisted to quantum phase slips.
However, the agreement with the theory is only qualitative, since the theoretical expectations for the exponent $\alpha$ have not been confirmed. In particular the experimental exponents $\alpha$ are interaction-independent and they are of the order of unity \cite{Tanzi16}, whereas the theoretical exponents depend on the interaction and they can be an order of magnitude larger than the measured exponents \cite{Danshita13}. This disagreement could be due to the range of velocities explored, much larger than in the theory, and to the lattice strength, much lower than in the theory \cite{Danshita13}. Nevertheless, new theoretical studies performed in the regime of shallow lattices have demonstrated that the experimentally observed $v$-independent regime is well described by a TAPS mechanism \cite{Kunimi}. By predicting the damping rate $G$ induced by thermally activated phase slips in the regime of low velocity and weak interaction, in fact, the theory shows a good agreement with the experimental values, thus suggesting the interpretation of thermal activation in the $v\ll v^*$ regime. Furthermore, the damping rates measured with large velocity and large interaction, whose values are definitely larger than the calculations with the TAPS assumption, strongly suggest a quantum activation of phase slips in this regime. The experimental results are thus compatible with the TAPS to QPS crossover. However, the quantitative comparison between theory and experiment in the QPS regime, necessary to demonstrate the nature of the observed crossover, is still missing.

\section{Conclusion}
Ultracold quantum gases are revealing themselves as a new platform for the investigation of the QPS phenomenon. They are allowing to explore transport phenomena in various configurations that are the analogous of established condensed-matter systems, such as superfluid rings, Josephson junctions and 1D wires. Thanks to their ample tunability and to the relative easiness of modelling, they might allow to investigate aspects of QPS that are not accessible in condensed matter. Seminal experiments in one-dimensional systems have shown strong dissipation dominated by thermal or quantum phase slips \cite{Fertig05,Kasevich,Ketterle07,Demarco08,Tanzi13,Tanzi16}, the onset of a temperature-independent regime of dissipation \cite{Demarco08}, and apparently also a crossover from thermal to quantum phase slips controlled by the velocity \cite{Tanzi16,Scaffidi}. An exhaustive picture of QPS in ultracold superfluids is however still missing and many fundamental questions are still open. In the future, it might be interesting to investigate different types of obstacles, such as disorder \cite{Khlebnikov} or isolated impurities \cite{Buchler}. Furthermore, it might be possible to observe directly individual QPS events or the interaction of few of them with interferometric techniques \cite{Hadzibabic,Hofferberth}, or by single atom detection \cite{Bakr,Sherson}.
\vskip6pt






\textbf{Acknowledgments;}
The authors acknowledge fruitfull discussions with I. Danshita and M. Kunimi.
This work was supported by the EC - H2020 research and innovation programme (Grant No. 641122 - QUIC) and by the Italian MIUR (Grant No. RBFR12NLNA - ArtiQuS)



\end{document}